\documentclass{elsart}
\def\cal{\mathcal}
\def\ba{\begin{eqnarray}}
\def\ea{\end{eqnarray}}
\def\ep{\varepsilon}
\usepackage{rotating}

\input{epsf.tex}

\usepackage{amssymb}

\begin{document}
\date{}
\begin{frontmatter}



\title{Lowest order QED radiative corrections to five-fold differential
cross section of hadron leptoproduction  
}


\author[Duke]{I. Akushevich}
\ead{igor.akushevich@duke.edu}
\author[Minsk]{A. Ilyichev}
\ead{ily@hep.by}
\author[Genoa,MSU]{M. Osipenko}
\ead{osipenko@ge.infn.it}

\address[Duke]{Duke University, Durham, USA}
\address[Minsk]{National Scientific
and Educational Center of Particle and High Energy Physics of the Belarusian State University,
220040  Minsk,  Belarus
}
\address[Genoa]{Istituto Nazionale di Fisica Nucleare, Sezione di Genova,
	16146 Genoa, Italy}
\address[MSU]{Skobeltsyn Institute of Nuclear Physics, 119992 Moscow, Russia}

\begin{abstract}
The contribution of exclusive radiative tail 
 to the cross section of
semi-inclusive hadron leptoproduction has been calculated
exactly for the first time. 
Although the experience of inclusive data analyses suggests us that 
the contribution of radiative tail from the elastic peak is of particular importance, 
similar effects in the semi-inclusive process
were only recently estimated in the peaking approximation.
The explicit expressions for the lepton part of the lowest order QED
contribution of exclusive radiative tail
to the five-fold differential cross section are obtained and
discussed. Numerical estimates, 
provided within
Jefferson Lab  kinematic conditions,  demonstrate rather large effects of the
exclusive radiative tail in the region 
at semi-inclusive threshold and for high energy of detected hadron. 
\end{abstract}

\begin{keyword}
Radiative corrections \sep semi-inclusive deep inelastic scattering
 \sep exclusive radiative tail
\PACS  13.40.Ks \sep 13.60.-r
\end{keyword}
\end{frontmatter}

\section{Introduction}
The semi-inclusive deep inelastic scattering of a lepton on the 
nucleon
represents an important tool for studying strong interaction. 
The possibility of representing a semi-inclusive hadron leptoproduction (SIHL) cross section as a convolution of
the virtual photon absorption 
by the quarks inside the nucleon and the subsequent quark
hadronization allows one to investigate these mechanisms separately. 
The SIHL  experiment provides not only complete
information on the longitudinal parton momentum distributions 
available in inclusive deep inelastic scattering (DIS)
experiments, but also an insight on the hadronization process
and on parton orbital momenta. 

\begin{figure}[t!]
\unitlength 1mm
\begin{tabular}{ccc}
\begin{picture}(80,80)
\put(-27,0){
\epsfxsize=9cm
\epsfysize=9cm
\epsfbox{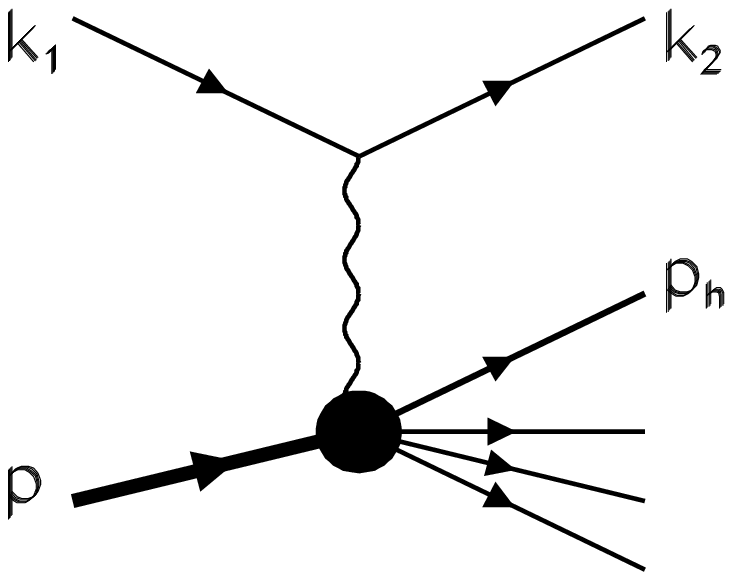}
\put(-50,38){\mbox{a)}}
}
\end{picture}
&
\begin{picture}(80,80)
\put(-60,0){
\epsfxsize=9cm
\epsfysize=9cm
\epsfbox{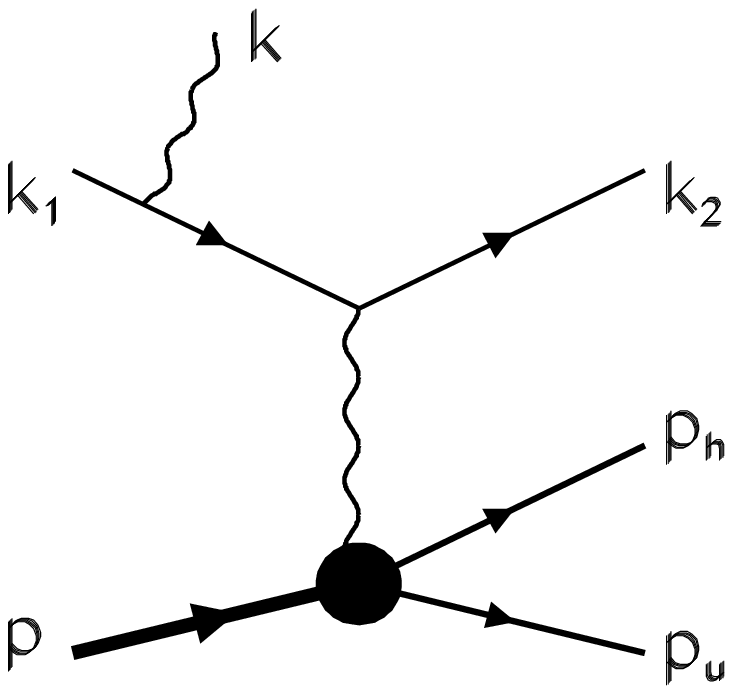}
\put(-50,38){\mbox{b)}}
}
\end{picture}
&
\begin{picture}(80,80)
\put(-93,0){
\epsfxsize=9cm
\epsfysize=9cm
\epsfbox{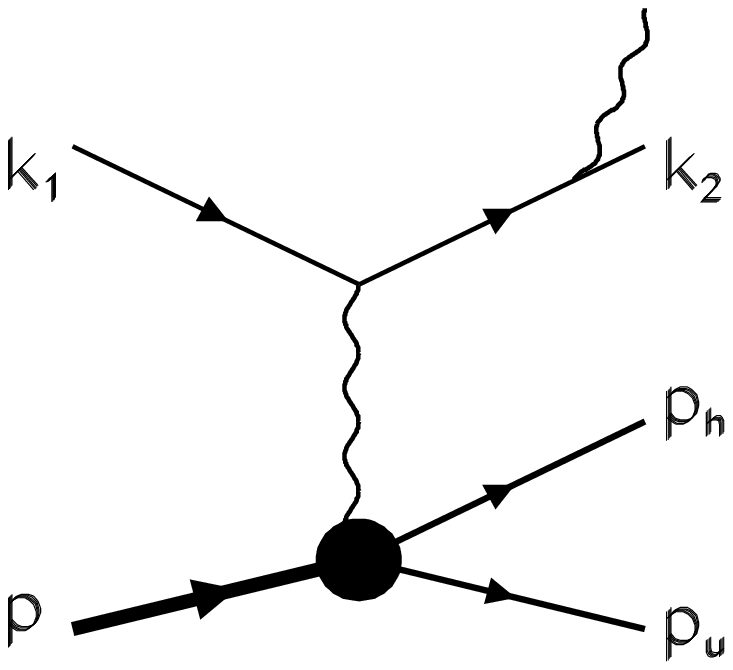}
\put(-50,38){\mbox{c)}}
}
\end{picture}
\end{tabular}
\vspace*{-40mm}
\caption{\label{gr}
\it Feynman graphs for the lowest order
(a) and the exclusive radiative tail  contributions to SIHL cross section
with real photon emission from the initial (b)
and final (c) leptons.
}
\end{figure}

It is well known that SIHL events are altered
by the real photon emission from the lepton and hadron legs
as well as by additional virtual particle contributions.
Due to i) the fact that most of the outgoing
particles in SIHL remain undetected
and ii) the finite resolution of experimental equipment,
not all events with the real photon emission
can be removed experimentally.
Moreover, the contribution of events with an additional exchange of virtual particles 
cannot be removed at all.
As a result the measured SIHL cross section
includes not only the lowest order contribution which is the process of 
interest (Fig.~\ref{gr}~(a)), but also the higher order  
effects
 whose contribution
has to be removed from the data. 
Since the latter cannot be extracted by 
experimental methods, the corresponding radiative corrections (RC) have to be
calculated theoretically.

The primary step in the solution of the task on RC calculation 
in the lepton nucleon scattering assumes the 
calculation of the part of the total lowest order QED correction 
that includes real photon emission
from lepton leg as well as the additional virtual photon between the initial and final leptons
and the correction due to virtual photon self-energy. 
There are two basic reasons for why other types of RC, such as box-type 
contribution or real photon emission from hadrons, are less 
important.
The first is that these corrections do not contain the
leading order contribution
which is proportional to the logarithm of the lepton mass, and therefore, their contribution
is much smaller comparing to  RC from lepton part. The second is that the
calculation of these effects  requires
additional assumptions about hadron interaction, so it has additional
pure theoretical uncertainties, which are hardly 
controlled. 

In the very first detailed SIHL experiments~\cite{Cornell} RC
were unknown and Monte Carlo simulations based on the approach
from Ref.~\cite{MoTsai} were used to correct the data.
The results of this Monte Carlo method however, have not been tested so far
against the direct calculations of radiative effects.
Meanwhile most experiments at high energies
\cite{HERMES} neglected RC completely \cite{EMC}. 

The calculations of the lepton part of the lowest order QED RC to 
SIHL cross sections were performed in Refs.~\cite{sorshum,hap} using
Bardin-Shumeiko covariant approach~\cite{BSh}. 
In Ref.~\cite{sorshum} the radiative effects were calculated
for the three-dimensional cross section of unpolarized and polarized SIHL
(target and lepton were longitudinally polarized)
and the FORTRAN code for numerical estimates was provided as a patch
(named SIRAD) to POLRAD code~\cite{POLRAD}.
In Ref.~\cite{hap} RC for the unpolarized 
five-fold differential cross section
have been computed and FORTRAN code HAPRAD has been developed.
However in both papers, RC do not include the contribution of the radiative tail
from the exclusive reaction at the threshold. In
inclusive DIS experiments analogous
effects from the elastic radiative tail \cite{MoTsai,ASh} 
give an important contribution to the observable cross section and, moreover,
there exist kinematic regions (e.g. at high $y$ or $Q^2$ and  small $x$),
where this contribution is dominant. This additional term of RC to
SIHL has been investigated until now only in the peaking approximation \cite{JLab_pub}.

In the present Letter the contribution of the lepton part of the lowest order
QED RC to SIHL due to exclusive radiative tail
is calculated exactly for the first time. This is done
using the approach from Ref.~\cite{diff,exc} and notations from Ref.~\cite{hap}.
The RC were calculated for complete five-fold differential cross section.
The technique of exact calculation of the lowest order RC (over $\alpha$)
is used in this Letter. 
The accuracy of the calculation is defined by accuracy of
numerical integration, which can be easily controlled.
Whereas actual values of the correction depend on 
the particular choice of the exclusive reaction parameters.

The rest of the Letter is organized as follows. In Section 2 
we define 
kinematics of the investigated processes, obtain explicit expressions 
for the contribution of exclusive radiative tail 
 to the five-fold differential cross section of
SIHL, and investigate its analytical 
properties considering the soft photon limit.
Discussion of the numerical results and concluding remarks are presented in Section 3.
Also the explicit expressions allowing for the presentation of the 
results in closed form are given in Appendix.


\section{Kinematics and explicit expressions}
Feynman graphs giving the Born as well as
the lepton part of lowest order QED correction to SIHL cross section from
the exclusive radiative tail
are shown on Fig.~\ref{gr}~(b,c). 
The radiative tail is generated by
the real photon emission from the lepton leg accompanying the exclusive
leptoproduction:
\ba
l(k_1)+p(p)\rightarrow l^\prime(k_2)+h(p_h)+u(p_u)+\gamma(k).
\ea
Following the notations of Ref.~\cite{hap} we call measured in the final state hadron $h$,
which is observed in coincidence with the scattered lepton $l^\prime$.
The second hadron $u$ that completes the exclusive reaction remains
undetected.
Here  $k_1$ ($k_2$) is the four-momentum of the initial (final) lepton
($k_1^2=k_2^2=m^2$),
$p$ is the target four-momentum,
($p^2=M^2$),
$p_h$ ($p_u$) is the four-momentum of the detected (undetected) hadron
($p_h^2=m_h^2$, $p_u^2=m_u^2$),
and $k$ is the emitted real photon four-momentum ($k^2=0$).

The set of variables describing the five-fold differential SIHL
cross section can be chosen as follows:
\ba
x=-\frac{q^2}{2qp},\;
y=\frac{qp}{k_1p},\;
z=\frac{p_hp}{pq},\;
t=(q-p_h)^2,\;
\phi_h,
\ea
where $q=k_1-k_2$ and $\phi_h$ is the angle between
$({\bf k_1},{\bf k_2})$ and $({\bf q},{\bf p_h})$
planes in the target rest frame  reference system (${\bf p}=0$).

For the description of the real photon emission 
we will use the following three variables:
\ba
R=2kp,\;\tau=\frac{2kq}R, \; \phi_k
\ea
with $\phi_k$ being the angle between
$({\bf k_1},{\bf k_2})$ and $({\bf q},{\bf k})$
planes in the target rest frame reference system.

We also will use the following Lorentz invariants:
\ba
&&S=2k_1p,\; X=2k_2p=(1-y)S,\;Q^2=-q^2=xyS,
\nonumber \\
[2mm]
&&W^2=S_x-Q^2+M^2,\; S_x=S-X,\; S_p=S+X,\; 
\lambda _q=S_x^2+4 M^2Q^2,
\nonumber \\
[1mm]
&&
\lambda _s=S^2-4 M^2m^2,\;
V_{1,2}=2k_{1,2}p_h=2(a^{1,2}+b \cos \phi _h),\;
\nonumber \\
[1mm]
&&
\mu=\frac{2kp_h}R=2(a^k+b^k\cos(\phi_k-\phi _h)),
\nonumber \\
&&f=\frac{2k(p+q-p_h)}R=1+\tau-\mu,
\label{inv}
\ea
where the explicit expressions for $a^{1,2,k}$, $b$ and $b^k$
coefficients can be found in Appendix B.

It is also useful to define
the non-invariant quantities  describing 
kinematics of 
the detected hadron  such as
its energy $E_h$, longitudinal $p_l$ and transvers $p_t$
three-momenta with respect to the virtual photon direction 
in the target rest frame.
This quantities can be expressed through
the Lorentz invariants introduced above in the following way:
\begin{eqnarray}
\label{ppt}
E_h&=&z\nu =\frac{zS_x}{2M},
\;
p_l=\frac{M}{\sqrt{\lambda_q}}(t-m_h^2+Q^2+2\nu E_h),
\;
p_t^2=E_h^2-p_l^2-m_h^2,
\nonumber\\
\nu&=&\frac {S_x}{2M},
\end{eqnarray}
where $\nu $ is a virtual photon energy
in the targets rest frame.
  
Instead of the commonly accepted
in SIHL analyses variable $p_t^2$
we will
use the  variable $t$.
This is dictated not only by the fact that $t$ is
Lorentz invariant, but also by 
the necessity to
distinguish the forward and backward hemispheres, mixed in the 
$p_t^2$-differential cross section. At intermediate energies of Jefferson Lab,
the contribution
of backward kinematics is significant, in particular for heavy hadrons
detected in the final state.
This backward kinematics is related to the target fragmentation mechanism
described in terms of fracture functions~\cite{Trentadue}.
Also one can notice that $p_t^2$-differential cross section is divergent
in the completely transverse case $p_l=0$ making difficult numerical 
integrations\footnote{See Fig.~\ref{pt2t} and comments after Eq.~\ref{tmin} for details.}.

According to Eq.~33 of Ref.~\cite{exc} the contribution
of the one-photon emission from the lepton leg to the
exclusive  hadron leptoproduction cross section
can be presented as the integral of the squared matrix elements 
described by Fig.~\ref{gr} (b,c) over the inelasticity
$v=(p+q-p_h)^2-m_u^2$ and the photon solid angle. 
The integration in our case is three-fold, because the
measured exclusive cross section is four-dimensional and the cross section with emission of one additional
photon is seven-dimensional. However, 
when we consider this contribution to the five-fold differential SIHL cross section, one photonic variable is fixed by
measurement, and the contribution has a form of a two-dimensional integral. 
Specifically, we use the inelasticity  $v$ as an observable in SIHL:
\ba
v&=&(1-z) S_x+t+M^2-m_u^2.
\label{vv}
\ea
At the same time the variable $R$ is fixed by both observable 
and two photonic variables in the following way:
\ba
R
&=&\frac{v}{f}
=\frac{(1-z) S_x+t+M^2-m_u^2}{1+\tau -2(a^k+b^k\cos(\phi_k-\phi _h))}.
\label{rv}
\ea
Hence the integration over other two unobserved photonic variables: $\tau $ and $\phi _k$ 
 requires for the calculation of the exclusive radiative tail contribution to SIHL.

The cross section responsible for the lepton part
of the exclusive radiative tail (see Fig.~\ref{gr}~(b,c))  
 is given 
by
\ba
d\sigma^R_{ex}&=&\frac{M^2_R}{2\sqrt{\lambda _s}
(2\pi)^8}\frac{d^3k_2}{2E_2}
\frac{d^3k}{2\omega}
\frac{d^3p_h}{2E_h}
\frac{d^3p_u}{2E_u}
\delta^4(p+q-p_h-k-p_u)
\nonumber \\
&=&\frac{M^2_R}{
(4\pi)^7}\frac{RSS_x^2}{f \lambda _s\lambda _q}
dxdydz dt d\phi _hd\tau d\phi _k. 
\label{main0}
\ea
The squared matrix element $M^2_R$ can be presented as a convolution 
of the leptonic and hadronic tensors. The former has well-known structure:
\ba
L_{\mu \nu}^R
&=&-\frac 12 {\rm Tr}[({\hat k}_2+m)\Gamma _{\mu \alpha }({\hat k}_1+m)
{\hat \Gamma }_{\alpha \nu }],
\nonumber \\
\Gamma _{\mu \alpha }&=&
\left (
\frac {k_{1\alpha }}{k k_1}-\frac {k_{2\alpha }}{k k_2}\right )\gamma_{\mu}
-\frac{\gamma _{\mu} {\hat k}\gamma _{\alpha}}{2kk_1}
-\frac{\gamma _{\alpha} {\hat k}\gamma _{\mu}}{2kk_2},\;
\nonumber \\
\hat {\Gamma }_{\alpha \nu }&=&
\left (
\frac {k_{1\alpha }}{k k_1}-\frac {k_{2\alpha }}{k k_2}\right )\gamma_{\nu}
-\frac{\gamma _{\alpha} {\hat k}\gamma _{\nu}}{2kk_1}
-\frac{\gamma _{\nu} {\hat k}\gamma _{\alpha}}{2kk_2}
,\;
\ea
while  the latter can be presented in a 
following covariant way:
\ba 
W^{\mu\nu}&=&-{\widetilde g}^{\; \mu\nu}{\cal H}_1
\;+\;{\widetilde p}^{\; \mu}{\widetilde p}^{\; \nu}{\cal H}_2
\;+\;{\widetilde p_h}^{\; \mu}{\widetilde p_h}^{\; \nu}{\cal H}_3
+ \;( {\widetilde p}^{\; \mu}{\widetilde p_h}^{\; \nu}
  +{\widetilde p_h}^{\; \mu}{\widetilde p}^{\; \nu}
     ){\cal H}_{4}
\label{hadt}
\ea
where
\ba
{\widetilde g}^{\; \mu\nu}=g^{\mu\nu}+\frac{q^{\mu}q^{\nu}}{Q^2}
,\quad
{\widetilde p}^{\; \mu}=p^{\mu}+\frac{q^{\mu}\;pq}{Q^2}
,\quad
{\widetilde p_h}^{\; \mu}=p_h^{\mu}+\frac{q^{\mu}\;p_hq}{Q^2},
\ea
and the Lorentz invariant structure functions ${\mathcal H}_i$
can be related to the exclusive photoabsorption cross sections as shown in Appendix A.
After convolution of the leptonic and hadronic tensors the squared matrix element reads:
\ba
M_R^2=\frac{(4\pi\alpha)^3}{{\widetilde Q}^4}L_{\mu \nu}^RW_{\mu \nu}
=-\frac{2(4\pi\alpha)^3}{{\widetilde Q}^4R}\sum_i\theta_i{\cal H}_i.
\label{mr2}
\ea

Combining Eqs. \ref{main0} and \ref{mr2} 
 we obtain the contribution of the exclusive radiative tail to
SIHL cross section: 
\ba
\frac {d\sigma_{ex}^R}{dxdydzdtd\phi_h}=
-\frac{\alpha ^3SS_x^2}{2^7\pi^4
\lambda _s
\lambda _q}
\int\limits_{\tau_{min}}^{\tau_{max}}d\tau
\int\limits_{0}^{2\pi}d\phi_k
	\sum\limits_{i=1}^4
\theta_i(\tau,\phi_k)
\frac {{\mathcal H}_i({\widetilde W^2},{\widetilde Q^2},
{\widetilde t})}{f{\widetilde Q}^4}.
\label{main}
\ea
The integration limits over $\tau$ are
given by $\tau_{min,max}=(S_x\pm \sqrt{\lambda _q})/2M^2$.
Quantities $\theta_i(\tau,\phi_k)$ have the following form:
\ba
\theta _i(\tau,\phi_k)=\frac{4F_{IR}\theta_{i}^B}R
+\theta_{i2}+R\theta_{i3},
\label{maint}
\ea
where $F_{IR}$, $\theta_{i2}$ and $\theta_{i3}$ are defined in Appendix B
and 
\ba
&\theta_{1}^B=Q^2-2m^2, \;\;\;\; \qquad \;
&\theta_{2}^B=(SX-M^2Q^2)/2, \nonumber \\
&\theta_{3}^B=(V_1V_2-m_h^2Q^2)/2,\;
&\theta_{4}^B=(V_2S+V_1X-z Q^2 S_x )/2 .
\label{coef}
\ea
The structure functions ${\mathcal H}_i$ depend on shifted kinematic variables
modified with respect to ordinary ones by the real photon emission:
\ba
{\widetilde W^2}&=&(p+q-k)^2=W^2-R(1+\tau),\;
\nonumber \\
{\widetilde Q^2}&=&-(q-k)^2=Q^2+R\tau,\;
\nonumber \\
{\widetilde t}&=&(q-p_h-k)^2=t+R-v.
\label{td}
\ea
\begin{figure}[t!]
\unitlength 1mm
\hspace*{-1.5cm}
\vspace*{5mm}
\begin{tabular}{cc}
\begin{picture}(80,80)
\put(10,0){
\epsfxsize=6.5cm
\epsfysize=6cm
\epsfbox{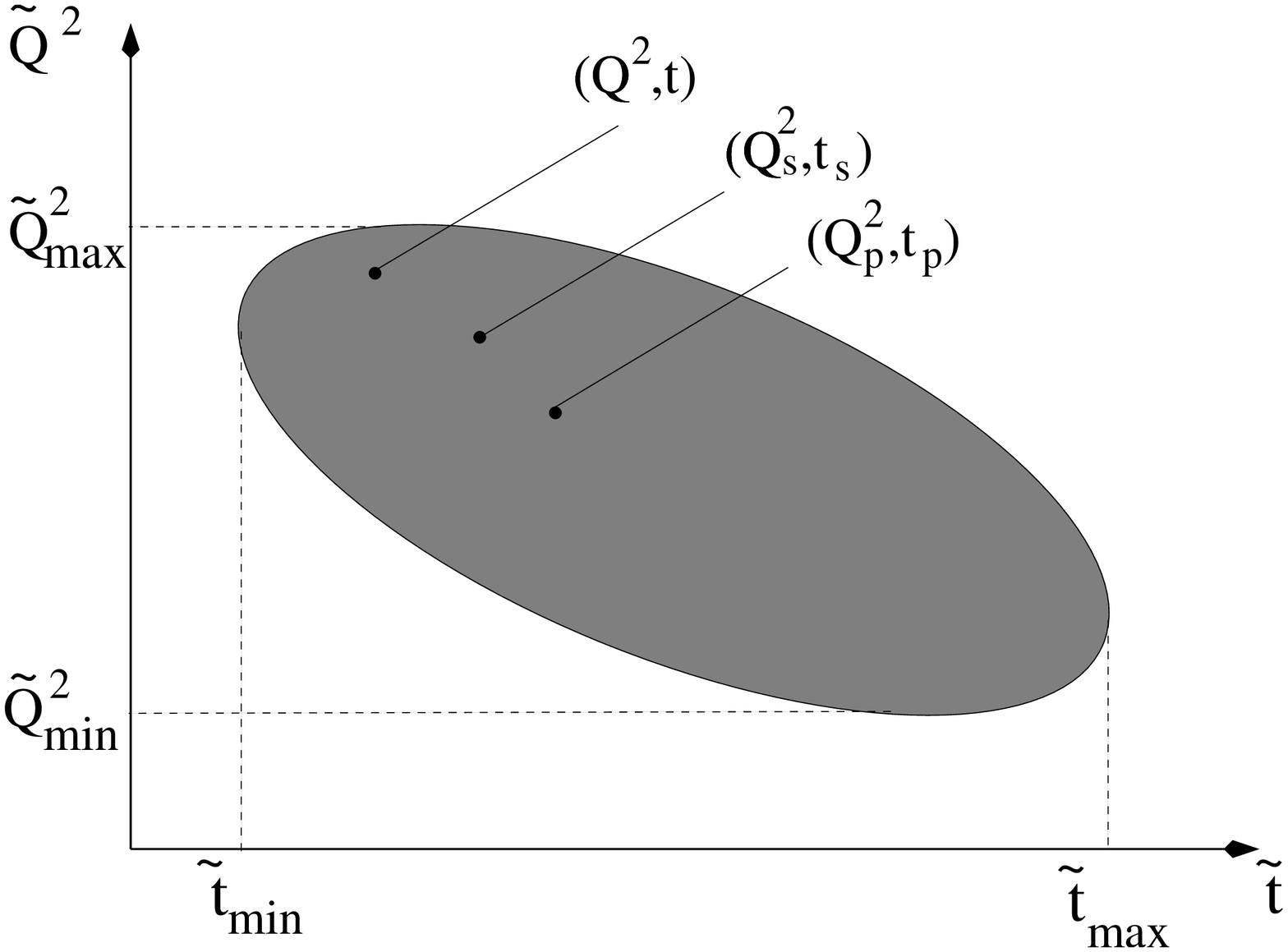}
}
\end{picture}
&
\begin{picture}(80,80)
\put(0,0){
\epsfxsize=6.5cm
\epsfysize=6cm
\epsfbox{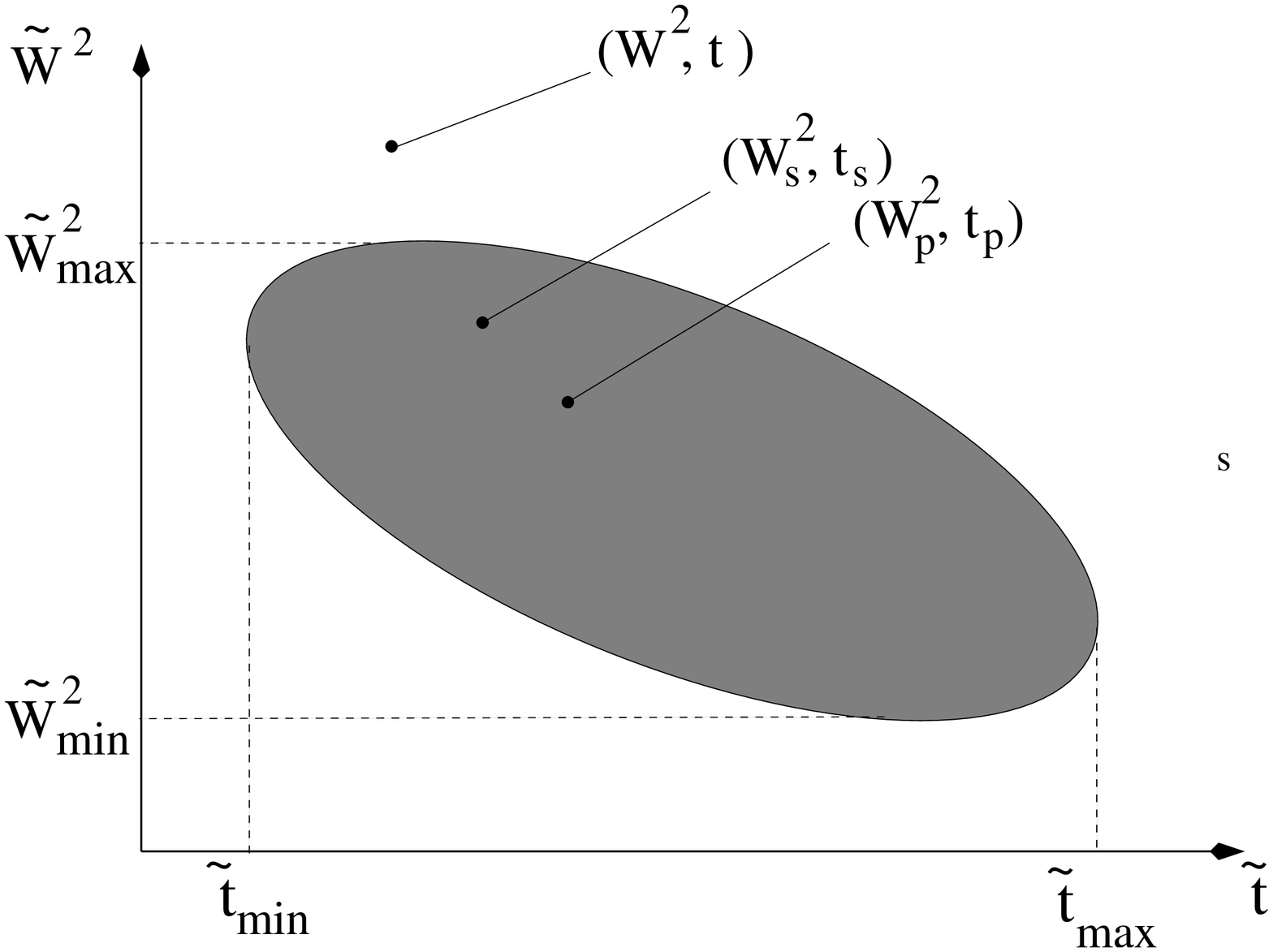}
}
\end{picture}
\end{tabular}
\vspace*{-10mm}
\caption{\label{td4} \it
The region of changes variables $\widetilde Q^2$,
$\widetilde W^2$ and $\widetilde t$:
Born point is defined as $(Q^2,t)$, $(W^2,t)$. 
Two points corresponding to the collinear photon emission along 
initial and final leptons are presented as $(Q^2_s,t_s)$, $(W^2_s,t_s)$
and $(Q^2_p,t_p)$, $(W^2_p,t_p)$ respectively.
}
\end{figure}
The region of changes for these variables is depicted in Fig.~\ref{td4}. 
Maximum and minimum values of these variables are defined in the following way:
\ba
{\widetilde W^2_{max/min}}&=&
W^2-\frac{v(C_{W^2}
\mp\sqrt{C_{W^2}^2-4W^2(v+m_u^2)})
}
{2(v+m_u^2)}
,
\nonumber \\
{\widetilde Q^2_{max/min}}&=&
Q^2-\frac{v(C_{Q^2}
\mp\sqrt{C_{Q^2}^2+4Q^2(v+m_u^2)})
}
{2(v+m_u^2)}
,
\nonumber \\
{\widetilde t_{max/min}}&=&
t-\frac{v(C_{t}
\mp\sqrt{C_{t}^2-4t(v+m_u^2)})
}
{2(v+m_u^2)}
,
\label{td2}
\ea
where 
$C_{W^2}=W^2+v-m_h^2+m_u^2$, $C_{Q^2}=Q^2-S_x-t+m_h^2$ and
$C_{t}=t+v-M^2+m_u^2$.

In this figure one can see the Born point as well as points that
correspond to the so-called collinear singularity (that was only used in \cite{JLab_pub} 
for peaking approximation): $\tau=\tau_s\equiv -Q^2/S$ ($\tau=\tau_p\equiv Q^2/X$), 
$\phi _k=0$, when the real photon is emitted along the momentum of the initial
(final) lepton. These points correspond to the following shifted variables:
\ba
W^2_{s,p}&=&W^2-R_{s,p}(1+\tau_{s,p}),\;
\nonumber \\
Q^2_{s,p}&=&Q^2+R_{s,p}\tau_{s,p},\;
\nonumber \\
t_{s,p}&=&t+R_{s,p}-v,
\label{sp1}
\ea
where
\ba
R_s&=&{vS\lambda_q \over 
\lambda_q(S-Q^2)+(SS_x+2Q^2M^2)t_q-zQ^2S_xS_p+2\sqrt{\Lambda}\cos (\phi_h)p_t},
\nonumber \\[0.3cm]
R_p&=&{vX\lambda_q \over 
\lambda_q(X+Q^2)+(XS_x-2Q^2M^2)t_q-zQ^2S_xS_p+2\sqrt{\Lambda}\cos (\phi_h)p_t},
\label{sp2}
\ea
$t_q=t+Q^2-m_h^2$ and $\Lambda=Q^2\lambda_q(SX-M^2Q^2)$. 

If in Eq.~\ref{main} we restrict our consideration 
only to the soft photon emission the result has to be 
proportional to the Born contribution to
the exclusive  cross section 
with a coefficient, independent of
type of considered process. To obtain this well-known  relation between
 contributions of the soft-photon emission and the Born to the cross section 
of the exclusive process, 
it is necessary to integrate  Eq.~\ref{main} over $z$ keeping only 
the photons with energy $\omega$  in the limits:
$\omega_{min}\leq \omega \leq \omega_{max}\ll all\; energies\; and\; masses$.
The corresponding Born contribution
 reproduces the cross section of the exclusive
leptoproduction and can be expressed in terms of the coefficients
(\ref{coef}) and the structure functions ${\mathcal H}_i$:
\ba
\frac {d\sigma^B_{ex}}{dxdydtd\phi_h}=
\frac{\alpha^2 SS_x}{16\pi^2Q^4\lambda _s\sqrt{\lambda _q}}
\sum\limits_{i=1}^4{\mathcal H}_i(W^2, Q^2,t)
\lim\limits_{z \to z_0}\theta_i^B .
\label{born}
\ea
The integration variable $z$ and the photon energy in the target rest frame are related as 
\ba
z=\frac{t+S_x+M^2-m_u^2-2fM\omega}{S_x},
\ea
while the limit $z_0$ corresponds to the situation when emitted photon energy is equal to zero:
$z_0=(t+S_x+M^2-m_u^2)/S_x$.

Therefore taking into account
\ba
\int \limits_{\tau _{min}}^{\tau _{max}}d\tau
\int \limits_{0}^{2\pi }d\phi_kF_{IR}
&=&-4\pi \sqrt{\lambda _q}
\left ( 
\frac {Q^2+2m^2}{\sqrt{\lambda _m}}
\log \frac {\sqrt{\lambda _m}+Q^2}{\sqrt{\lambda _m}-Q^2}-1 \right ) ,
\ea
where $\lambda _m=Q^2(Q^2+4m^2)$, finally we obtain the sought equality in the form: 
\ba
\frac{d\sigma _{ex}^{soft}}
{dxdydtd\phi _h}
&=&\frac{2\alpha}
{\pi}
\log\frac {\omega_{max}}{\omega_{min}}
\Biggl ( 
\frac {Q^2+2m^2}{\sqrt{\lambda _m}}
\log \frac {\sqrt{\lambda _m}+Q^2}{\sqrt{\lambda _m}-Q^2}
-1 \Biggr ) 
\nonumber \\[2mm] &&\qquad\qquad\qquad\times
\frac {d\sigma^B_{ex}}{dxdydtd\phi_h},
\ea
or, in the limit $m\to 0$,
\ba
\frac{d\sigma _{ex}^{soft}}
{dxdydtd\phi _h}
=\frac{2\alpha}
{\pi}
\log\frac {\omega_{max}}{\omega_{min}}
\left ( \log \frac {Q^2}{m^2}-1 \right )
\frac {d\sigma^B_{ex}}{dxdydtd\phi_h}.
\ea

Thus, we reproduced expected result for the cross section of soft photon
irradiation (e.g., see Eq. (7.64) of \cite{999}).

\section{Discussion of Numerical Results and Concluding Remarks}

In this Section the contribution of the exclusive radiative tail is illustrated in several examples 
investigated under kinematical conditions of the current experiments on the SIHL measurements. 
For this purpose the FORTRAN code was developed\footnote{This code is available at
http://www.hep.by/RC.}. 

Most recent experiments measuring SIHL are being performed at Jefferson Lab.
In particular, the large acceptance of CLAS detector allows for extraction of
the information about the five-fold differential SIHL cross section in a rather
wide kinematic region that covers almost the whole $z$-range as well as
the entire $\phi_h$-range.   
In this section we present numerical results for the
five-fold differential SIHL cross section in CLAS kinematic conditions.  

\begin{figure}[t!]
\unitlength 1mm
\vspace*{4mm}
\begin{picture}(80,80)
\put(25,0){
\epsfxsize=9cm
\epsfysize=9cm
\epsfbox{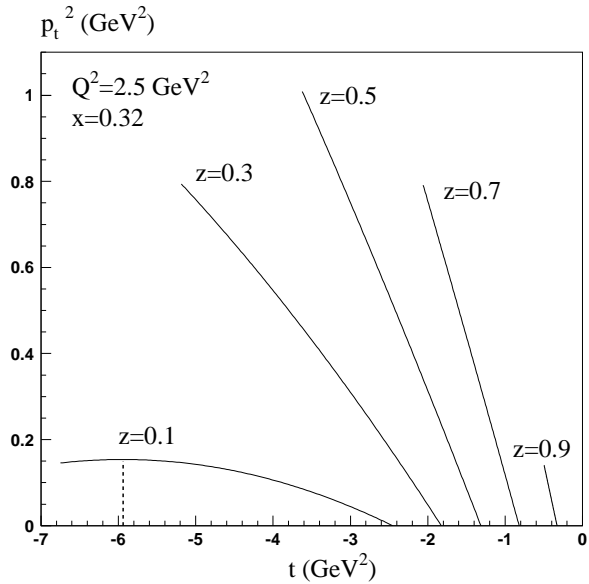}
}
\end{picture}
\vspace*{-5mm}
\caption{\label{pt2t} \it
$t$-dependence of $p_t^2$ at different $z$. The dashed line shows 
$t$-points where $p_l$ changes sign.
}
\end{figure}

One remark has to be discussed before presenting numerical estimates. Consider 
$t$-dependence of $p_t^2$ at fixed $Q^2$, $x$ and different $z$
that is presented in Fig. \ref{pt2t}
for $\pi^+$-electroproduction in electron-proton scattering. 
The upper $t$-limit corresponds to the maximum value of the 
detected hadron longitudinal momentum $p_l$:
\ba
t_{max}=m_h^2-Q^2+\frac 1{2M^2}(\sqrt{\lambda _q(z^2S_x^2-4 M^2m_h^2)}-zS_x^2),
\label{tmax}
\ea
while the lowest one at low energy is given by the SIHL threshold
\ba
t_{min}=(m_u+m_{\pi })^2-M^2-(1-z)S_x,
\label{tmin}
\ea
i.e. when missing mass square
\ba
M^2_X=(p+q-p_h)^2=(1-z)S_x+t+M^2
\label{mm}
\ea
reaches its minimal value. Here $m_{\pi }$ is a pion mass.
A remarkable feature of this plot is that the curve $z=0.1$ crosses the 
point where $p_l$ 
changes the sign and both positive and negative values of $p_l$ give the same $p_t$.
As it was mentioned above, due to the common
denominator $|p_l|$ the $p_t^2$-differential SIHL cross section diverges
at this point.

To estimate the value of RC we introduce the radiative correction
factor in the standard way
\ba 
\delta=
\frac{\sigma_{obs}}{\sigma_{B}},
\label{delta}
\ea 
where $\sigma_{obs}$ ($\sigma_{B}$) is the radiatively corrected (Born)
five-fold differential cross section of the semi-inclusive hadron leptoproduction.

The analytical expressions of RC obtained in previous section can be applied to leptoproduction of any hadrons
observable in the lepton-nucleon scattering. 
However we restrict our numerical studies to the case of $\pi^+$ production in electron-proton scattering. 
The calculation of RC factor requires applying the parameterization of the
photoabsorption cross sections. We use the model developed by collaboration 
{ MAID 2003}~\cite{maid}. This model provides parameterizations for each of the required  
photoabsorption cross sections which are continuous in whole kinematic region.  
It   accurately predicts the behavior of the
photoabsorption cross sections in the resonance region and has true asymptotic behavior for
higher $W$ and $Q^2$ by means of the fit from Ref.~\cite{Browman}.
The numerical estimation of this effect requires knowledge of the structure functions within the kinematical restriction for the
shifted variables presented in Fig.~\ref{td4}.  However the most important region is concentrated near the 
$s$- and $p$-collinear singularity (see Eq.~\ref{sp1}) where the integrand expression reaches its maximum value. 
A possible effect of the specific choice of photoabsorption cross sections (e.g., choice of the 
{ MAID 2003} model) can be investigated by comparing the predictions of the model with experimental data or other model predictions in this particular region. For example, the CLAS  kinematics restrictions (i.e. $E_{beam}=6$ GeV, $1$ GeV$^2<Q^2< 7$ GeV$^2$,
$0<p_t<1.5$ GeV) for collinear singularity region are
$0.07$~GeV$^2<Q^2_{s,p}<10$ GeV$^2$, $1.17$~GeV$^2<W^2_{s,p}<10$ GeV$^2$ and
$-8$~GeV$^2<t_{s,p}<8\cdot 10^{-3}$ GeV$^2$. Since the { MAID 2003}  describes experimental data in this region sufficiently 
well \cite{maid} (at least for $d\sigma_L/d\Omega_h$ and $d\sigma_T/d\Omega_h $ 
provided the main contribution for the total cross section) and provides convenient parametric 
form for all required photoproduction cross sections, the choice of the { MAID 2003} seems to be reasonable and practical. 
In application of data analyses collected in the specific regions especially in situations when the other two 
photoabsorption cross sections $d\sigma_{TT}/d\Omega_h$ and $d\sigma_{LT}/d\Omega_h $ 
give rather large contribution (e.g., for measurements of $t$- or $\phi_h$-dependence), 
the model independence has to always be tested by comparison with data (or other models) in the collinear regions. 
This is especially important in the light of recent investigations which demonstrated that in certain cases the 
{ MAID 2003} 
can be imperfect \cite{EX}.

\begin{figure}[t!]
\unitlength 1mm
\hspace*{-1.5cm}
\vspace*{5mm}
\begin{tabular}{cc}
\begin{picture}(80,80)
\put(5,0){
\epsfxsize=8cm
\epsfysize=8cm
\epsfbox{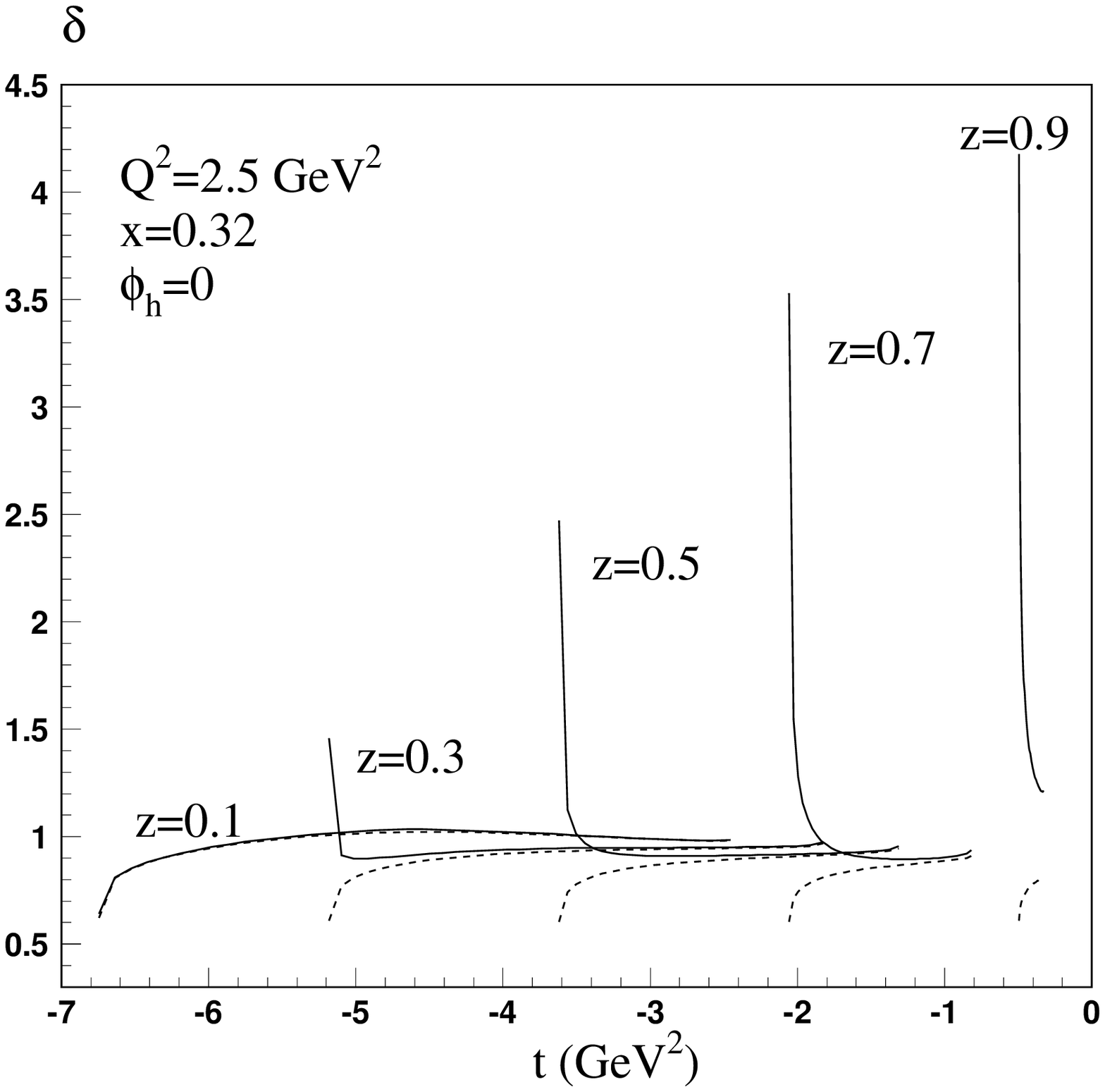}
}
\end{picture}
&
\begin{picture}(80,80)
\put(-10,0){
\epsfxsize=8cm
\epsfysize=8cm
\epsfbox{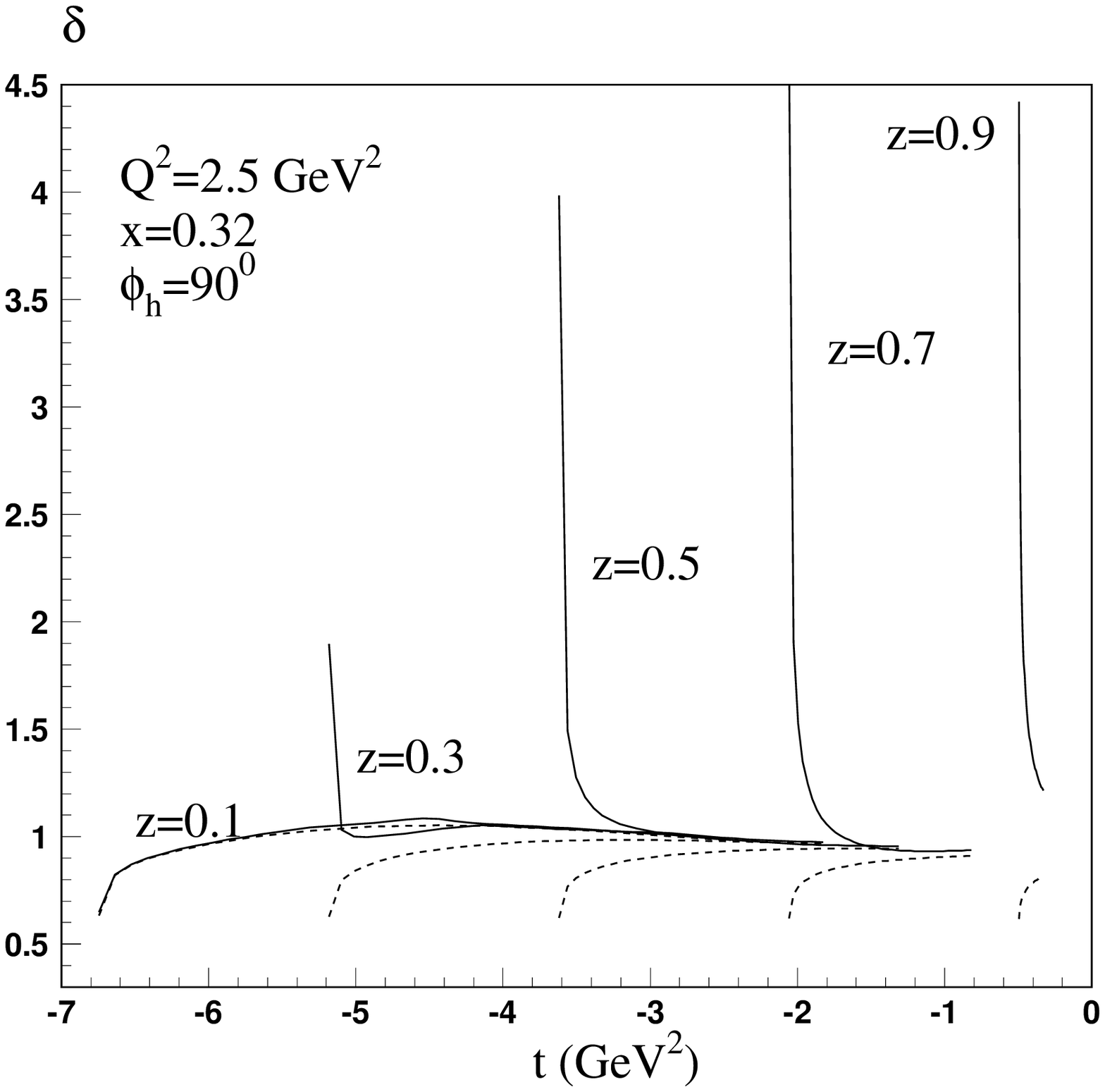}
}
\end{picture}
\\[-12mm]
\begin{picture}(80,80)
\put(35,0){
\epsfxsize=8cm
\epsfysize=8cm
\epsfbox{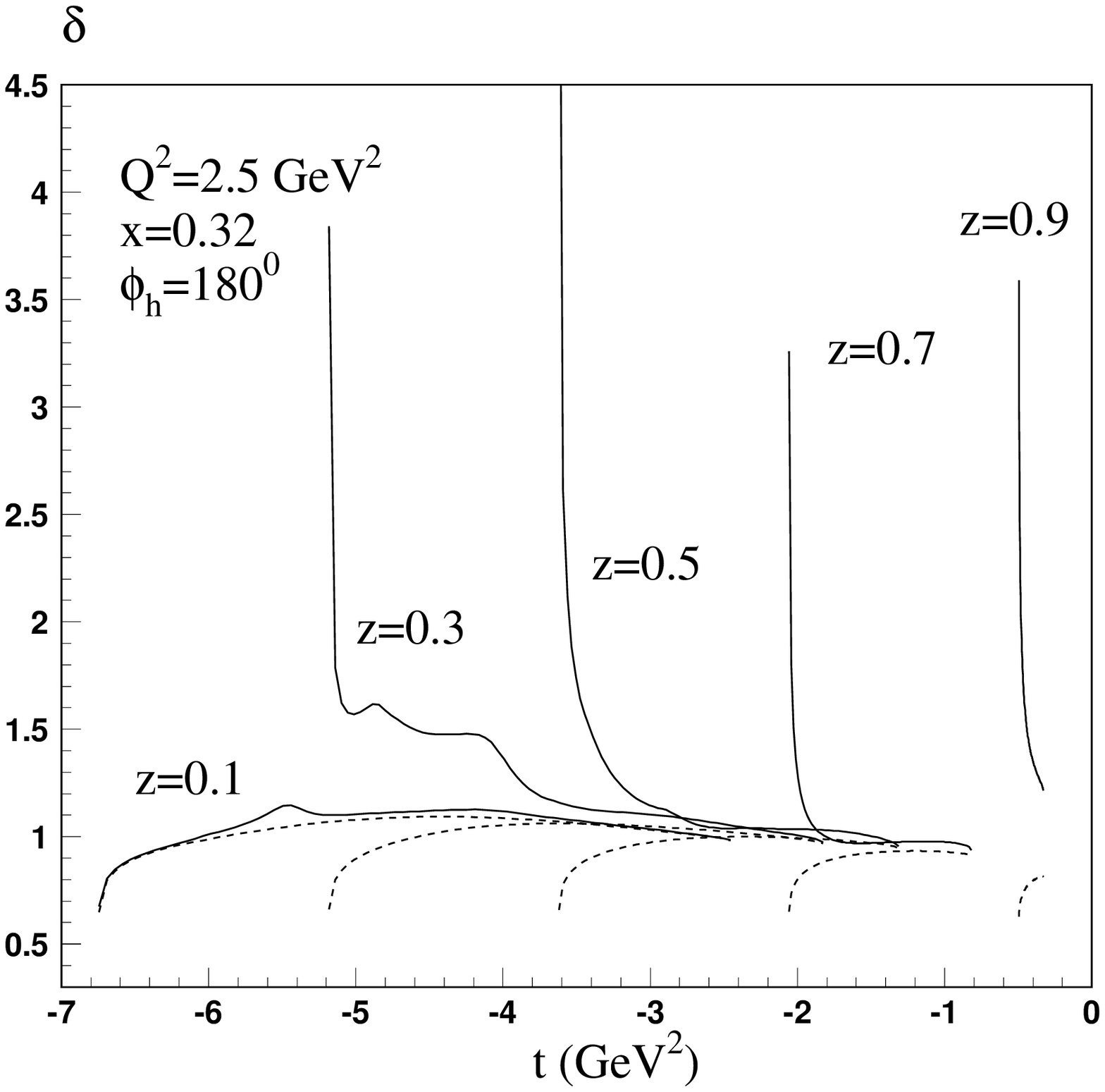}
}
\end{picture}
\end{tabular}
\vspace*{-10mm}
\caption{\label{ect} \it
$t$-dependence of the RC factor (\ref{delta}) for the semi-inclusive $\pi^+$ electroproduction
at fixed proton for lepton beam energy 6 GeV:
solid lines show the total correction,
dashed lines represent the correction excluding the exclusive radiative tail
calculated in this Letter.}
\end{figure}

\begin{figure}[t!]
\unitlength 1mm
\vspace*{5mm}
\begin{picture}(80,80)
\put(25,0){
\epsfxsize=9cm
\epsfysize=9cm
\epsfbox{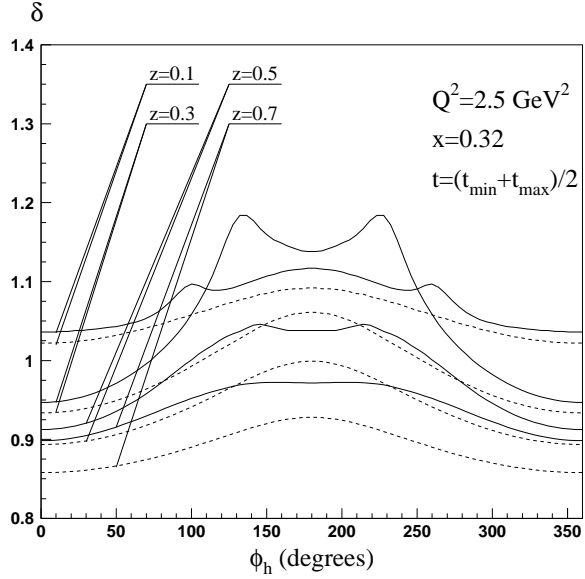}
}
\end{picture}
\vspace*{-5mm}
\caption{\label{ecfi}  \it
$\phi_h$-dependence of the RC factor (\ref{delta}) for the semi-inclusive $\pi^+$ electroproduction
at fixed proton for lepton beam energy 6 GeV:
solid lines show the total correction,
dashed lines represent the correction excluding the exclusive radiative tail
calculated in this Letter.}
\end{figure}

\begin{figure}[t!]
\unitlength 1mm
\hspace*{-1.5cm}
\vspace*{5mm}
\begin{tabular}{cc}
\begin{picture}(80,80)
\put(5,0){
\epsfxsize=8cm
\epsfysize=8cm
\epsfbox{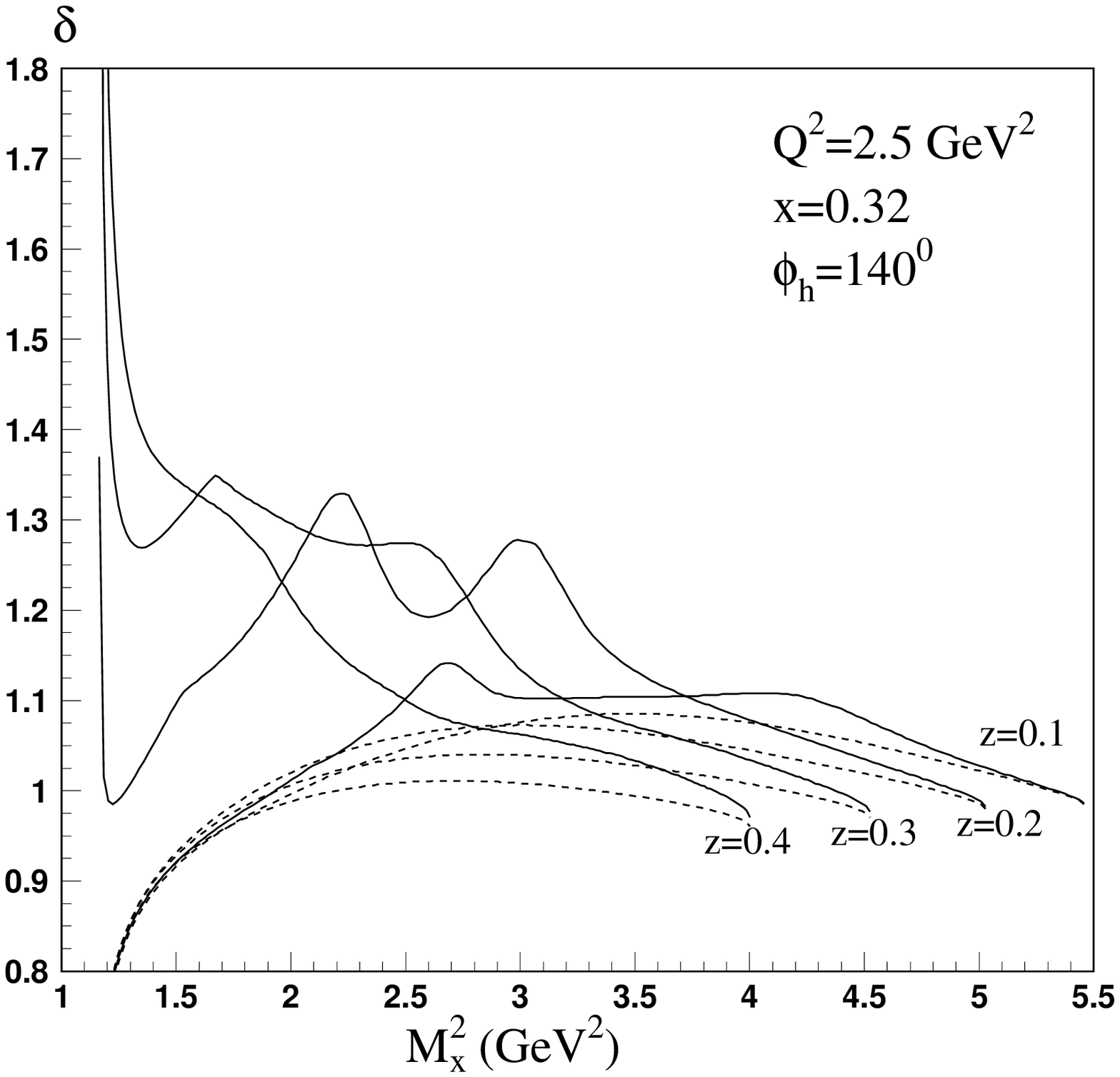}
}
\end{picture}
&
\begin{picture}(80,80)
\put(-10,0){
\epsfxsize=8cm
\epsfysize=8cm
\epsfbox{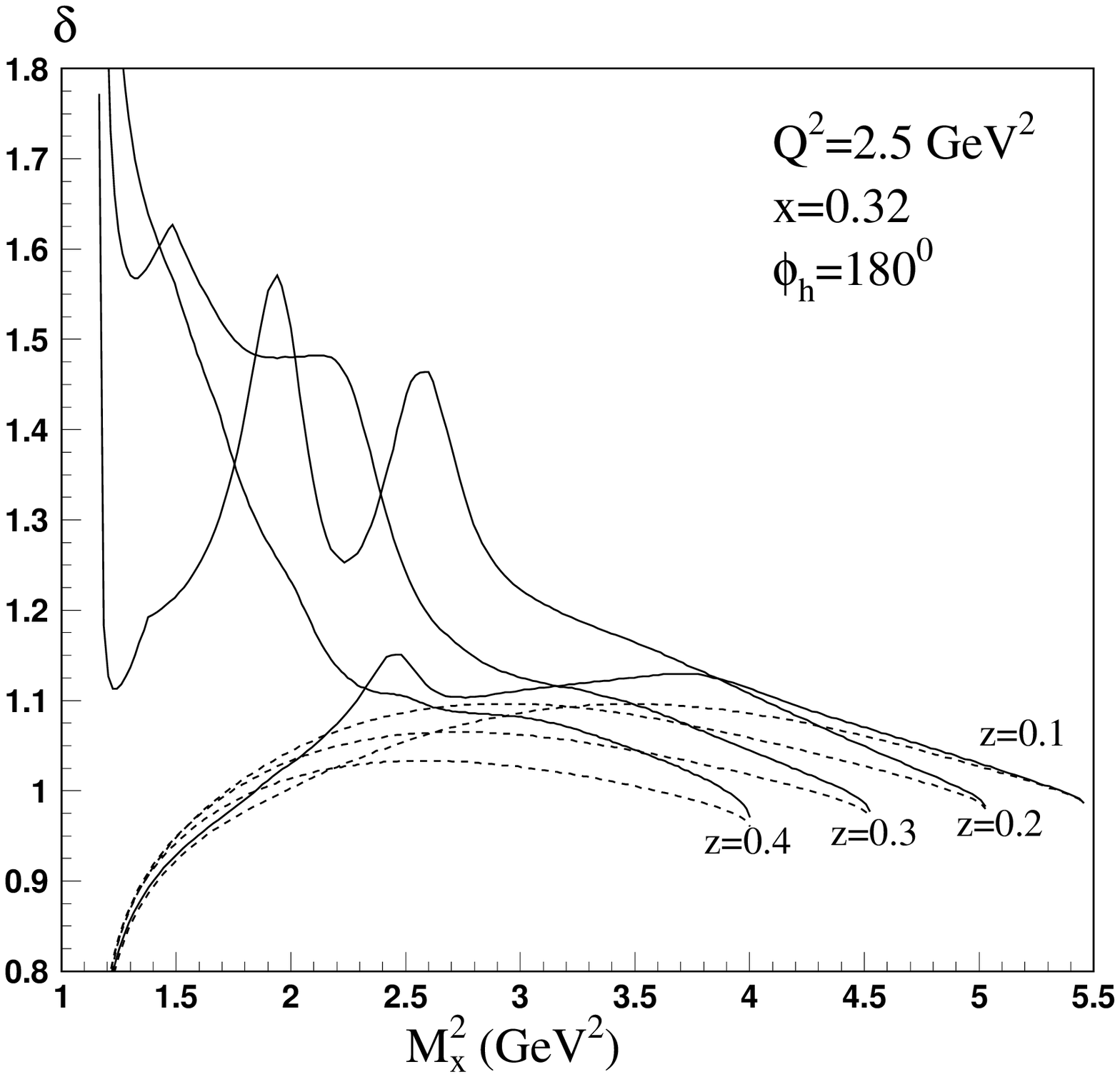}
}
\end{picture}
\end{tabular}
\vspace*{-10mm}
\caption{\label{ecm2} \it
$M_X^2$-dependence of the RC factor (\ref{delta}) for the semi-inclusive $\pi^+$ electroproduction
at fixed proton for lepton beam energy 6 GeV:
solid lines show the total correction,
dashed lines represent the correction excluding the exclusive radiative tail
calculated in this Letter.}
\end{figure}

Examples of RC factor including the exclusive radiative tail contribution
are shown on Figs.~\ref{ect}, \ref{ecfi} and \ref{ecm2}. 

It could be seen in Fig.~\ref{ect} that 
the exclusive radiative tail contribution at $z \sim 0.1$ is small or even negligible
but rapidly increases with the growing $z$ near SIHL  threshold, i.e., when $t \to t_{min}$. 
Such behavior appears from the first term in Eq.~\ref{maint} due to smallness
of the denominator in expression for $R$ defined by Eq.~\ref{rv}. 
In contrast to the elastic
radiative tail contribution to inclusive DIS, 
the minimally allowed value of ${\widetilde Q^2}$ (see Eq.~\ref{main}) 
in the integrand of Eq.~\ref{main} does not reach the region close to the photon point region 
${\widetilde Q^2}\to 0$ where the exclusive cross section increases rapidly.
As a result, the so-called $t$-peak (or Compton peak) often essentially contributing to the cross section 
of the elastic radiative tail \cite{AKP,AISh} does not appear in the considered case, and the main contribution 
to exclusive radiative tail appears from the collinear region.

The absolute value
of the exclusive radiative tail rapidly increases with growing the invariant $t$
(or missing mass of the detected lepton-hadron system). However, the SIHL
cross section increases with $t$ much faster making the relative
contribution of the exclusive radiative tail small or negligible at large $t$.
Meanwhile, the situation
changes to the opposite at small $t$ i.e. close to the threshold where the exclusive radiative tail exceeds
the SIHL cross section (Fig.~\ref{ect}). 

Moreover,
one can see in Figs.~\ref{ect} and \ref{ecfi} that the contribution of the exclusive radiative tail can 
significantly modify the
 $\phi_h$-distributions at middle $t$ distorting usual
$A+B\cos{\phi _h}+C\cos{2\phi _h}$ behavior.

Fig.~\ref{ect} and \ref{ecfi} illustrate that if the at low $\phi _h$ the exclusive radiative tail
contribution to SIHL reaches rather high values near the pion threshold, then 
with rising $\phi _h$ up to 180$^0$ the correction is rather large for much wider kinematical region. 
The behavior of the RC factor as a function of missing mass square (\ref{mm})
at $\phi _h=140^0$ and $\phi _h=180^0$ is
shown in Figs.~\ref{ecm2}. From this plot it can be seen
that the exclusive radiative tail has a significant contribution 
in $M_X^2>(m_u+m_{\pi })^2$ already at $z=0.2$ that shifts to the pion threshold
with the growing $z$.
 Particularly at $z=0.2$ and $\phi _h=140^0$
the contribution of the exclusive radiative tail to SIHL
exceed 20\% in the  $M_X^2\approx 3$ GeV$^2$ region.

The bump structures in the exclusive tail contribution seen 
in Figs.~\ref{ect} and \ref{ecm2} are due
to the contribution of nucleon resonances in the MAID parameterization.

In experimental data analyses the five-fold cross sections are estimated in specific bins. 
There exist several schemes of how radiative correction can be applied to the cross 
section observed in a certain bin. The simplest and most practical variant is to 
calculate the RC for the center of the bin, i.e., for means of kinematic variables defining the bin. 
If the bin is broad over one or several kinematic variables then the Born cross section and 
the contribution of the exclusive radiative tail can change differently over the bin. 
In this case the integration over the bin has to be performed 
with taking into account all experimental cuts which are used 
by experimentalists to form the bin. The exact procedure to apply the 
RC to data collected in such bins supposes integration of the cross section of 
RC over a bin. Since this procedure is computationally extensive, often approximate procedures are used.   
One possible scheme of such approximate integration is a so-called 'event-by-event' scheme where 
reweighing the RC factor defined by Eq.~\ref{delta} is applied for each reconstructed event.
Some of experimental cuts used by experimentalists can essentially influence the RC factor for the bin.
One often used cut is the cut on missing mass or inelasticity. This cut allows to avoid a contribution of resonances 
and also it is important for RC calculation. 
Tab.~\ref{tabq1} presents the results of RC calculation for these two approaches to binning forming (i.e., with and without cutting the resonance region) as 
used in ref \cite{avakian}. As one can see, applying the cut on missing mass allows to suppress the 
contribution of the exclusive radiative tail. Note, there are experimental situations where RC including the contribution from the resonance region are of great importance. Examples include the analysis of the measurements 
of the  threshold reactions, e.g., ``quark-hadron duality'' which is based on 
a comparison of production in the resonance region with 
extrapolation of DIS measurements, or reanalysis of older Cornell data from 
\cite{Cornell} which were collected without any cuts of the threshold region. Therefore, our program includes an option to apply a cut on missing mass squared
at the integrand of the cross section of the exclusive radiative tail.
\begin{table}[!t]
\begin{tabular}{|c|c|c|c|c|c|c|c|c|}
\hline
&$x$&$Q^2$&$z$&$p_t$&
\multicolumn{4}{c|}{$\delta-1 $}
\\
\cline{6-9} 
&&GeV$^2$&&GeV&
$\phi_h=0$&
$\phi_h=\pi  /4$&
$\phi_h=3\pi /4$&
$\phi_h=\pi$
\\ \hline
&0.18&1.10&0.61&0.46&0.01558&0.01558&0.01559&0.01561\\ \cline{2-9} 
&0.24&1.30&0.61&0.42&0.02373&0.02374&0.02375&0.02378\\ \cline{2-9} 
&0.31&1.60&0.61&0.41&0.03692&0.03693&0.03695&0.03699\\ \cline{2-9} 
&0.37&2.00&0.61&0.39&0.04650&0.04651&0.04653&0.04656\\ \cline{2-9} 
&0.27&1.46&0.54&0.43&0.01753&0.01753&0.01755&0.01757\\ \cline{2-9} 
\rule{2mm}{0pt}
\begin{rotate}{90}
$M_X>1.1$ GeV
\end{rotate}
&0.27&1.44&0.61&0.43&0.02869&0.02870&0.02872&0.02875\\ \cline{2-9} 
&0.27&1.44&0.69&0.42&0.05847&0.05848&0.05851&0.05857\\ \cline{2-9} 
&0.27&1.43&0.77&0.36&0.1508 &0.1508 &0.1509 &0.1510\\  \hline
&0.18&1.10&0.58&0.43&0.01286&0.01286&0.01287&0.01289\\ \cline{2-9} 
&0.24&1.40&0.57&0.36&0.01458&0.01459&0.01459&0.01461\\ \cline{2-9} 
&0.31&1.70&0.56&0.32&0.02241&0.02241&0.02242&0.02244\\ \cline{2-9} 
&0.37&2.00&0.56&0.28&0.03573&0.03574&0.03575&0.03577\\ \cline{2-9} 
\rule{2mm}{0pt}
\begin{rotate}{90}
$M_X>1.4$ GeV
\end{rotate}
&0.26&1.44&0.54&0.38&0.01505&0.01481&0.01483&0.01484\\ \cline{2-9} 
&0.25&1.41&0.61&0.34&0.02130&0.02072&0.02073&0.02075\\ \cline{2-9} 
&0.23&1.37&0.69&0.30&0.03438&0.03315&0.03316&0.03318\\ \cline{2-9} 
&0.20&1.26&0.77&0.24&0.05859&0.05602&0.05604&0.05606\\ \hline
\end{tabular}
\\
\caption{
Relative exclusive radiative tail contribution at different $\phi _h$
to the observed cross section for the kinematical points of ref.
\cite{avakian}}
\label{tabq1} 
 \end{table}
	
Summarizing, the exclusive radiative tail contribution to complete five-fold differential
unpolarized SIHL cross section has been calculated exactly
for the first time. Respective FORTRAN code for
the numerical estimation is opened for the scientific community.
Numerical analysis performed for kinematical conditions of the current experiments 
at JLab demonstrated that the RC to the SIHL coming from the exclusive radiative tail
is high in the regions of small $t$ and close to the threshold while 
for $\phi _h \sim $ 180$^0$ the kinematical region where RC is 
important is 
much wider.
This contribution significantly modifies $\phi_h$-asymmetries of the SIHL cross section.
The present approach is quite general and can be extended to other
SIHL reactions providing knowledge of the exclusive
cross section at the threshold.

The calculated correction to the SIHL due to the radiative tail from exclusive processes 
is important and its contribution always has to be taken into account in analyses of data 
in current and future experiments on the SIHL. Currently, this correction is ignored or analyzed 
in the peaking approximation \cite{JLab_pub},  quality of which can be evaluated only by comparison 
with exact formulae presented in this Letter. Several sources of systematical uncertainties have 
to be investigated in the data analyses including i) the specific choice of the model for 
the photoproduction cross sections, ii) quality of peaking approximation, if 
this approximation is used instead of the exact formulae, 
and iii) the choice of specific scheme of radiation correction procedure 
in a specific bin if it is used instead of exact integration of the radiative tail cross section over the bin.      

\vspace*{5mm}
{\bf Acknowledgments.} One of us (A.I.) would like to thanks the staff of
Istituto Nazionale di Fisica Nucleare (Genova, Italy) for their generous hospitality 
during his visit.

\appendix*

\section*{Appendix A. Structure functions}
\renewcommand{\theequation}{A.\arabic{equation}}
The following expressions relate the structure functions 
incoming into
(\ref{hadt}) to the Born photoabsorption cross sections, 
\begin{eqnarray}\label{SFs}
{\cal H}_1(W^2,Q^2,t)&=&C
\left (\frac{d\sigma_T}{d\Omega _h}-\frac{d\sigma_{TT}}{d\Omega _h}\right ),
\nonumber\\
{\cal H}_2(W^2,Q^2,t)&=&{2C\over\lambda_q}
\biggl[2Q^2\biggl(\frac{d\sigma_T}{d\Omega _h}-\frac{d\sigma_{TT}}{d\Omega _h}+\frac{d\sigma_{L}}{d\Omega _h}\biggr)
-2TQ\frac{d\sigma_{LT}}{d\Omega _h}
\nonumber\\&&
+T^2\frac{d\sigma_{TT}}{d\Omega _h}\biggr],
\nonumber\\
{\cal H}_3(W^2,Q^2,t)&=&{2C\lambda_q \over \lambda_l} \frac{d\sigma_{TT}}{d\Omega _h},
\nonumber\\
{\cal H}_4(W^2,Q^2,t)&=&{2C\over
\sqrt{\lambda_l}}\left(T\frac{d\sigma_{TT}}{d\Omega _h}-Q\frac{d\sigma_{LT}}{d\Omega _h}\right),
\end{eqnarray}
in such a way, that five-fold differential cross section for exclusive 
leptoproduction has a standard form \cite{maid,maid2}
\ba
\frac{d\sigma}{dE_2d\Omega_2d\Omega_h}&=&
\Gamma
\Biggl[
\frac{d\sigma_T}{d\Omega _h}
+\ep\frac{d\sigma_L}{d\Omega _h}
+\sqrt{2\ep(1+\ep)}\frac{d\sigma_{LT}}{d\Omega _h}\cos \phi _h
\nonumber \\&&
+\ep\frac{d\sigma_{TT}}{d\Omega _h}\cos 2\phi _h
\Biggr],
\nonumber \\
\Gamma&=&\frac {\alpha }{2\pi ^2}\frac {E_2 }{E_1}
\frac {\kappa_{\gamma} }{Q^2}
\frac 1{1-\ep},\;
\kappa_{\gamma}=\frac {W^2-M^2}{M^2},\;
\nonumber \\
\ep&=&
\Biggl(1+\frac{\lambda _q}{2(S X-M^2Q^2)}\Biggr)^{-1}.
\ea
Here $E_1=S/2M$ ($E_2=X/2M$) is the initial (final) lepton energy in the target rest 
frame system, $d\Omega_2$ ($d\Omega_h$) is a element of solid angle of scattering lepton
(detected hadron) in the target rest frame (c.m. system of the virtual photon and target),
\begin{eqnarray}
C&=&{16\pi(W^2-M^2)W^2\over \alpha
\sqrt{(W^2+m^2_h-m^2_u)^2-4m^2_hW^2}},
\nonumber\\
T&=&{S_x(t_q-2zQ^2) \over \sqrt{\lambda_l}},
\nonumber\\
\lambda_l&=&zS_x^2(zQ^2-t_q)-M^2t_q^2-m_h^2\lambda_q,
\nonumber\\
t_q&=&t+Q^2-m_h^2
\end{eqnarray}
and $Q=\sqrt{Q^2}$.
	
\section*{Appendix B. Some kinematic quantities}
\setcounter{equation}{0}
\renewcommand{\theequation}{B.\arabic{equation}}
The scalar products of $p_h$ in $V_{1,2}$ and $\mu$ 
(see Eq.~\ref{inv}) are expressed via coefficients
$a^1$, $a^2$, $b$,
$a^k$ and
$b^k$:
\begin{eqnarray}\label{a065}
2Ma^1&=&SE_h-(SS_x+2M^2Q^2)p_l\lambda_q^{-1/2},
\nonumber\\
2Ma^2&=&XE_h-(XS_x-2M^2Q^2)p_l\lambda_q^{-1/2},
\nonumber\\
b&=&-p_t\sqrt{\lambda/\lambda_q},
\nonumber\\
2Ma^k&=&E_h-p_l(S_x-2M^2\tau)\lambda_q^{-1/2},
\nonumber\\
b^k&=&-Mp_t\sqrt{\lambda_{\tau}/\lambda_q},
\end{eqnarray}
where
\ba 
\lambda_{\tau}=(\tau-\tau_{min})(\tau_{max}-\tau),\;
\lambda=SXQ^2-M^2Q^4-m^2\lambda _q.
\ea 

Quantities $\theta_{ij}$ (see Eq.~\ref{maint}) have the following form:
\ba
\theta_{12}&=&4 F_{IR}\tau,
\nonumber \\
[2mm]
\theta_{13}&=&-4-2F_d\tau ^2,
\nonumber \\
[2mm]
2\theta_{22}&=&F_{1+}S_xS_p-F_d\tau S_p^2+2 m^2F_{2-}S_p
+2 F_{IR}(S_x-2 M^2\tau),
\nonumber \\
[2mm]
2\theta_{23}&=&F_d(4m^2+\tau (2M^2\tau-S_x))-F_{1+}S_p+4 M^2,
\nonumber \\
[2mm]
\theta_{32}&=&2(F_{IR}(\mu V_{-}-m_h^2\tau)+ m^2F_{2-}\mu V_{+}
+F_{1+}V_{+}V_{-}-F_d\tau V_{+}^2),
\nonumber \\
[2mm]
\theta_{33}&=&F_d(2 m^2 \mu^2+\tau (m_h^2\tau-\mu V_{-}))-2F_{1+}\mu V_{+}
+2 m_h^2,
\nonumber \\
[2mm]
\theta_{42}&=&-2F_d\tau V_{+}S_p+F_{1+}( S_xV_{+}+S_pV_{-})
\nonumber \\[2mm]
&&
+m^2F_{2-}(\mu S_p+2V_{+})+F_{IR}((\mu-2\tau z)S_x+2V_{-}),
\nonumber \\
[2mm]
2\theta_{43}&=&F_d(8\mu m^2+\tau ((2\tau z-\mu)S_x-2V_{-}))
-F_{1+}(\mu S_p+2V_{+})
\nonumber \\[2mm]
&&
+4S_xz.
\ea
Here $V_{\pm}=(V_1\pm V_2)/2$, $F_{IR}=m^2F_{2+}-(Q^2+2m^2)F_d$, and 
\ba
F_d=\frac 1{z _1z _2},\;
F_{1+}=\frac 1{z _1}+\frac 1{z _2},\;
F_{2-}=\frac 1{z _2^2}-\frac 1{z _1^2},\;
F_{2+}=\frac 1{z_2^2}+\frac 1{z _1^2}.
\ea
The variable $z_{1,2}$ can be expressed as in Ref.~\cite{hap}
\ba 
z_1&=&\frac {2kk_1}R=
\frac 1 {\lambda _q}
\left[Q^2S_p+\tau (S S_x+2M^2Q^2)
-2M\cos \phi_k\sqrt{\lambda_{\tau}\lambda}\right],
\nonumber \\
z_2&=&\frac {2kk_2}R=
\frac 1 {\lambda _q}
\left[Q^2S_p+\tau (X S_x-2M^2Q^2)
-2M\cos \phi_k\sqrt{\lambda_{\tau}\lambda}\right].
\ea

\end{document}